\documentclass{article}

\usepackage{microtype}
\usepackage{graphicx}
\usepackage{subfigure}
\usepackage{booktabs} 

\usepackage{hyperref}

\usepackage[accepted]{icml2022}

\usepackage{amsmath}
\usepackage{amssymb}
\usepackage{mathtools}
\usepackage{amsthm}

\usepackage[capitalize,noabbrev]{cleveref}

\theoremstyle{plain}
\newtheorem{theorem}{Theorem}[section]

\newtheorem{lemma}[theorem]{Lemma}

\theoremstyle{definition}
\newtheorem{definition}[theorem]{Definition}

\theoremstyle{remark}

\usepackage{dirtytalk}

\usepackage{mathtools}

\usepackage{multicol,lipsum}

\newcommand{\BEAS}{\begin{eqnarray*}}
	\newcommand{\EEAS}{\end{eqnarray*}}
\newcommand{\BEA}{\begin{eqnarray}}
\newcommand{\EEA}{\end{eqnarray}}
\newcommand{\mb}{\mathbb}
\newcommand{\mc}{\mathcal}
\newcommand{\BEQ}{\begin{equation}}
\newcommand{\EEQ}{\end{equation}}
\newcommand{\BIT}{\begin{itemize}}
	\newcommand{\EIT}{\end{itemize}}
\newcommand{\BNUM}{\begin{enumerate}}
	\newcommand{\ENUM}{\end{enumerate}}

	\newcommand{\D}{\mathcal{D}}
	
		\newcommand{\F}{\mathcal{F}}
				\newcommand{\G}{\mathcal{G}}
	\newcommand{\Pm}{\mathcal{P}}
	\newcommand{\te}{\theta}
	\newcommand{\mr}{\mathrm}
	\newcommand{\R}{\mathbb{R}}
		\newcommand{\W}{\mathcal{W}}
		\newcommand{\E}{\mathbb{E}}

 \newcommand{\M}{\mathcal{M}}

	\newcommand{\ve}{\varepsilon}
\newcommand{\BA}{\begin{array}}
	\newcommand{\EA}{\end{array}}

\newtheorem{myh}{A.}

\begin{document}
\icmltitlerunning{Stochastic Primal-Dual Deep Unrolling}

\twocolumn[
\icmltitle{Stochastic Primal-Dual Deep Unrolling}




\begin{icmlauthorlist}
\icmlauthor{Junqi Tang}{a}
\icmlauthor{Subhadip Mukherjee}{a}
\icmlauthor{Carola-Bibiane Sch\"onlieb}{a}

\end{icmlauthorlist}

\icmlaffiliation{a}{Department of Applied Mathematics and Theoretical Physics (DAMTP), University of Cambridge, UK}

\icmlcorrespondingauthor{Junqi Tang}{jt814@cam.ac.uk}

\icmlkeywords{Machine Learning, ICML}

\vskip 0.3in
]



\printAffiliationsAndNotice{} 

\begin{abstract}

We propose a new type of efficient deep-unrolling networks for solving imaging inverse problems. Conventional deep-unrolling methods require full forward operator and its adjoint across each layer, and hence can be significantly more expensive computationally as compared with other end-to-end methods that are based on post-processing of model-based reconstructions, especially for 3D image reconstruction tasks. We develop a stochastic (ordered-subsets) variant of the classical learned primal-dual (LPD), which is a state-of-the-art unrolling network for tomographic image reconstruction. The proposed learned stochastic primal-dual (LSPD) network only uses subsets of the forward and adjoint operators and offers considerable computational efficiency. We provide theoretical analysis of a special case of our LSPD framework, suggesting that it has the potential to achieve image reconstruction quality competitive with the full-batch LPD while requiring only a fraction of the computation. The numerical results for two different X-ray computed tomography (CT) imaging tasks (namely, low-dose and sparse-view CT) corroborate this theoretical finding, demonstrating the promise of LSPD networks for 
large-scale imaging problems. 
\end{abstract}

\section{Introduction}

Stochastic first-order optimization methods have become the de-facto techniques in modern data science and machine learning with exceptionally wide applications \cite{kingma2014adam,johnson2013accelerating,allen2017katyusha, chambolle2018stochastic}, due to their remarkable scalability to the size of the optimization problems. The underlying optimization tasks in many applications nowadays are large-scale and high-dimensional by nature, as a consequence of big-data and overparameterized models (for example the deep neural networks).

While well-designed optimization algorithms can enable efficient machine learning, one can, on the other hand, utilize machine learning to develop problem-adapted optimization algorithms via the so-called \say{learning-to-learn} philosophy \cite{andrychowicz2016learning, li2019learning}. Traditionally, the optimization algorithms are designed in a hand-crafting manner, with human-designed choices of rules for computing gradient estimates, step-sizes, etc, for some general class of problems. Noting that although the traditional field of optimization has already obtain lower-bound matching (aka, \say{optimal}) algorithms \cite{woodworth2016tight,lan2012optimal,lan2015optimal} for many important general classes of problems, for specific instances there could be still much room for improvement. For example, a classical way of solving imaging inverse problems would be via minimizing a total-variation regularized least-squares \citep{chambolle2016introduction} with specific measurement operators, which is a very narrow sub-class of the general class smooth and convex programs where these \say{optimal} optimization algorithms are developed for. To obtain optimal algorithm adapted for a specific instance of a class, the hand-crafted mathematical design could be totally infeasible, and very often we do not even know a tight lower-bound of it.

One of the highly active areas in modern data science is computational imaging (which is also recognized as low-level computer vision), especially medical imaging including X-ray computed tomography (CT) \citep{buzug2011computed}, magnetic resonance imaging (MRI) \citep{vlaardingerbroek2013magnetic}, and positron emission tomography (PET) \citep{ollinger1997positron}. In such applications, the clinics seek to infer images of patients' inner body from the noisy measurements collected from the imaging devices. Traditionally, stochastic optimization schemes have been widely applied in solving large-scale imaging problems due to their scalability\citep{kim2015combining,sun2019online}. Inspired by their successes, in our work, we focus on developing efficient learned stochastic algorithms tailored for solving imaging inverse problems.

\subsection{Contributions of this work}

In this work, we make five main contributions:

\begin{itemize}
    \item \textbf{Novel deep unrolling networks -- leveraging the machinery of modern stochastic optimization}
    
    For the first time, we propose a class of deep unrolling networks based on the principles of modern stochastic first-order optimization, for efficiently solving imaging inverse problems. We develop Learned Stochastic Primal-Dual (LSPD) network, and its variant LSPD-VR which is further empowered with stochastic variance-reduction trick \cite{johnson2013accelerating,defazio2014saga,chambolle2018stochastic}. Our proposed networks can be viewed as a minibatch extension of the state-of-the-art unrolling network -- Learned Primal-Dual (LPD) of \citet{adler2018learned}.
    
    \item \textbf{Motivational analysis of LSPD network}
    
    We provide a motivational analysis of a simple instance of our LSPD network, from the view-point of stochastic non-convex composite optimization. We provide upper and lower bounds of the estimation error under standard assumptions, suggesting that our proposed networks have the potential to achieve the same estimation accuracy as its full-batch counterpart.
    
    \item \textbf{Less is more -- the numerical effectiveness of LSPD in tomographic medical imaging}
    
    We numerically evaluate the performance of our proposed networks on two typical tomographic medical imaging tasks -- low-dose and sparse-view X-ray CT. We compare our LSPD and LSPD-VR with the full batch LPD. We found out that our networks achieves competitive image reconstruction accuracy with the LPD, while only requiring a fraction of the computation of it as a free-lunch, in both supervised and unsupervised training settings
    
    \item \textbf{Learning to image without ground-truth -- fully unsupervised training for unrolling networks}

    We provide the first result in the literature for fully unsupervised training of unrolling networks without any use of ground-truth data, utilizing the equivariance structure of the measurement systems \cite{chen2021equivariant}. This is highly desirable in clinical practice since the ground-truth data is usually expensive and idealistic. We present this numerical study in the Appendix.
    \item \textbf{Out of training distribution? Get boosted! \\-- Efficient instance-adaptation for LSPD in out-of-distribution reconstruction tasks.}
    
    In computational imaging practice, we often encounter scenarios where the measurement data is observed under slightly different noise distribution and/or modality compared to the data used for training the unrolling network. We propose an instance-adaptation framework to fine-tune the unrolling network and adjust it to the out-of-distribution data. Due to the efficiency of LSPD, We find numerically that our approach provides superior performance on instance-adaptation tasks for out-of-distribution reconstruction.

\end{itemize}

\section{Algorithmic Framework}

\subsection{Background}
In imaging, the measurement systems can be generally expressed as:
\begin{equation}\label{model}
    b = A x^\dagger + w,
\end{equation}
where $x^\dagger \in \mb{R}^d$ denotes the ground truth image (vectorized), and $A \in \mb{R}^{n \times d}$ denotes the forward measurement operator, $w \in \mb{R}^n$ the measurement noise, while $b \in \mb{R}^n$ denotes the measurement data. A classical way to obtain a reasonably good estimate of $x^\dagger$ is to solve a composite optimization problem:
\begin{equation}\label{1st-stage}
    x^\star \in \arg\min_{x \in \mb{R}^d} f_b(Ax) + r(x),
\end{equation}
where data fidelity term $f_b(Ax)$ is typically a convex function (one example would be the least-squares $\|b - Ax\|_2^2$), while $r(x)$ being a regularization term, for example the total-variation (TV) semi-norm, or a learned regularization \citep{mukherjee2021end, mukherjee2020learned}. A classical way of solving the composite optimization problem (\ref{1st-stage}) is via the proximal gradient methods \citep{chambolle2016introduction}, which are based on iterations of gradient descent step on $f$, proximal step on $r$ and momentum step using previous iterates for fast convergence \citep{beck2009fast,nesterov2007gradient}. 

Since modern imaging problems are often in huge-scales, deterministic methods can be very computationally costly since they need to apply the full forward and adjoint operation in each iteration. For scalability, stochastic gradient methods \citep{robbins1951stochastic} and ordered-subset methods \citep{erdogan1999ordered,kim2015combining} are widely applied in real world iterative reconstruction. More recent advanced stochastic variance-reduced gradient methods \citep{xiao2014proximal,defazio2014saga, allen2017katyusha,tang2018rest,driggs2020spring} have also been adopted in some suitable scenarios in imaging for better efficiency \citep{tang2019limitation,tang2020practicality, karimi2016hybrid}.

More recently, deep learning approaches have been adapted in imaging inverse problems, starting from the work of \cite{jin2017deep} on the FBP-ConvNet approach for tomographic reconstruction, and DnCNN \citep{zhang2017beyond} for image denoising. Remarkably, the learned primal-dual (LPD) network \citep{adler2018learned} which mimics the update rule of primal-dual gradient method and utilizes the forward operator and its adjoint within a deep convolutional network, achieves state-of-the-art results and outperforms primal-only unrolling approaches. Despite the excellent performance, the computation of the learned primal-dual method is significantly larger than direct approaches such as FBP-ConvNet.

\subsection{Stochastic primal-dual unrolling}
Now we start by presenting the motivation of our unrolling network, starting from the basic primal-dual gradient-based optimization algorithm. It is well-known that, if the loss function $f_b(\cdot)$ is convex and lower semi-continuous, we can reformulate the original objective function (\ref{1st-stage}) to a saddle-point problem:
\begin{equation}\label{saddle}
    [x^\star, y^\star] = \min_{x} \max_{y} \{r(x) + \langle Ax, y \rangle - f_b^*(y)\},
\end{equation}
where $f_b^*(\cdot)$ is the Fenchel conjugate of $f_b(\cdot)$:
\begin{equation}
    f_b^*(y) := \sup_{h} \{\langle h, y \rangle - f_b(h)\}
\end{equation}
The saddle-point problem (\ref{saddle}) can be efficiently solved by the primal-dual hybrid gradient (PDHG) method \citep{chambolle2011first}, which is also known as the Chambolle-Pock algorithm in the optimization literature. The PDHG method for solving the saddle-point problem obeys the following updating rule:
 \begin{eqnarray*}
 && \mathrm{\textbf{Primal-Dual Hybrid Gradient (PDHG)}} \\&& - \mathrm{Initialize}\ x_0, \Bar{x}_0 \in \mb{R}^d \ y_0 \in \mb{R}^p\\
 &&\mathrm{For} \ \ \ k = 0, 1, 2,...,  K\\
&&\left\lfloor
\begin{array}{l}
y_{k+1} = \mathrm{prox}_{\sigma f_b^*} (y_k + \sigma A \Bar{x}_k);\\
x_{k+1} = \mathrm{prox}_{\tau r} (x_k - \tau A^T y_{k+1});\\
\Bar{x}_{k + 1} = x_{k+1} + \beta (x_{k+1} - x_{k});
\end{array}
\right.
 \end{eqnarray*}
The PDHG algorithm takes alternatively the gradients regarding the primal variable $x$ and dual variable $y$ and performs the updates. In practice, it is often more desirable to reformulate the primal problem (\ref{1st-stage}) to the primal-dual form (\ref{saddle}), especially when the loss function $f$ is non-smooth (or when the Lipschitz constant of the gradient is large).

Currently most successful deep networks in imaging would be the unrolling schemes \citep{gregor2010learning} inspired by the gradient-based optimization algorithms leveraging the knowledge of the physical models. The state-of-the-art unrolling scheme -- learned primal-dual network of \citep{adler2018learned} is based on unfolding the iteration of PDHG by replacing the proximal operators $\mathrm{prox}_{\sigma f^\star}(\cdot)$ and $\mathrm{prox}_{\tau g}(\cdot)$ with multilayer convolutional neural networks $\Pm_{\te_p}(\cdot)$ and $\D_{\te_d}(\cdot)$, with sets of parameters $\te_p$ and $\te_d$,  applied on the both primal and dual spaces. The step sizes at each steps are also set to be trainable. The learned primal-dual with $K$ iterations can be written as the following, where the learnable paramters are $\{\te_p^k, \te_d^k, \tau_k, \sigma_k\}_{k=0}^{K-1}$:
 \begin{eqnarray*}
 && \mathrm{\textbf{Learned Primal-Dual (LPD)}} \\&&- \mathrm{Initialize}\ x_0 \in \mb{R}^d \ y_0 \in \mb{R}^p\\
 &&\mathrm{For} \ \ \ k = 0, 1, 2,...,  K-1\\
&&\left\lfloor
\begin{array}{l}
y_{k+1} = \D_{\te_d^k} (y_k, \sigma_k, A {x}_k , b);\\
x_{k+1} = \Pm_{\te_p^k}(x_k, \tau_k, A^T y_{k+1});\\
\end{array}
\right.
 \end{eqnarray*}

When the primal and the dual CNNs are kept fixed across the layers of LPD, it has the potential to learn both the data-fidelity and the regularizer (albeit one might need additional constraints on the CNNs to ensure that they are valid proximal operators). This makes the LPD parametrization more powerful than a learned proximal-gradient network (with only a primal CNN), which can only learn the regularization functional. The capability of learning the data-fidelity term can be particularly useful when the noise distribution is unknown and one does not have a clear analytical choice for the fidelity term.

In our new approach, we propose to replace the forward and adjoint operators in the full-batch LPD network of \cite{adler2018learned}, with only subsets of it. The proposed network can be view as an unrolled version of stochastic PDHG \citep{chambolle2018stochastic} (but with ordered-subsets, and without variance-reduction). We partition the forward and adjoint operators into $m$ subsets, and also the corresponding measurement data. In each layer, we use only one of the subsets, in a cycling order. Let ${\bf{S}} := [S_0, S_1, S_2,..., S_{m-1}]$ be the set of subsampling operators, then the saddle-point problem (\ref{saddle}) can be rewritten as:
\begin{equation}\label{saddle1}
    [x^\star, y^\star] = \min_{x} \max_{y} \{r(x) + \sum_{i = 0}^{m-1}\langle S_iAx, y_i \rangle - f_{b_i}^*(y_i)\}.
\end{equation}
Utilizing this finite-sum structure, our learned stochastic primal-dual (LSPD) network can be described as\footnote{Alternatively, one may also consider an optional learned momentum acceleration by keeping the memory of the outputs of a number of previous layers: $x_{k+1} = \Pm_{\te_p^k}(X_k, \tau_k, (S_iA)^T y_{k+1})$ where $X_k = [x_k, x_{k-1},.. x_{k-M}]$, at the cost of extra computation and memory. For such case the input channel of the subnets would be $M+1$.}:
 \begin{eqnarray*}
 && \mathrm{\textbf{Learned Stochastic Primal-Dual (LSPD)}} \\&&- \mathrm{Initialize}\ x_0 \in \mb{R}^d \ y_0 \in \mb{R}^{p/m}\\
 &&\mathrm{For} \ \ \ k = 0, 1, 2,...,  K-1\\
&&\left\lfloor
\begin{array}{l}
i = \mod(k,m); \\\mathrm{(or\ pick\ i\ from\ [0, m-1]\ uniformly\ at\ random)}\\
y_{k+1} = \D_{\te_d^k} (y_k, \sigma_k, (S_iA) {x}_k , S_ib);\\
x_{k+1} = \Pm_{\te_p^k}(x_k, \tau_k, (S_iA)^T y_{k+1});\\
\end{array}
\right.
 \end{eqnarray*}
In the scenarios where the forward operator dominates the computation in the unrolling network, for the same number of layers, our LSPD network is approximately $m$-time more efficient than the full-batch LPD network in terms of computational complexity. The LSPD we presented here describes a framework of deep learning based methods depending the parameterization of the primal and dual subnetworks and how they are trained. In practice the LPD and LSPD networks usually achieves best performance when trained completely end-to-end. While being the most recommended in practice, when trained end-to-end, it is almost impossible to provide any non-trivial theoretical guarantees. An alternative approach is to restrict the subnetworks across layers to be the same and train the subnetwork to perform denoising \citep{kamilov2017plug, sun2019online,tang2020fast, ono2017primal}, artifact removal \citep{liu2020rare}, or approximate projection to a image manifold \citep{rick2017one}, leading to a plug-and-play \citep{venkatakrishnan2013plug, romano2017little, reehorst2018regularization} type of approach with theoretical convergence guarantees.

\begin{figure}[t]
   \centering

    {\includegraphics[width= .45\textwidth]{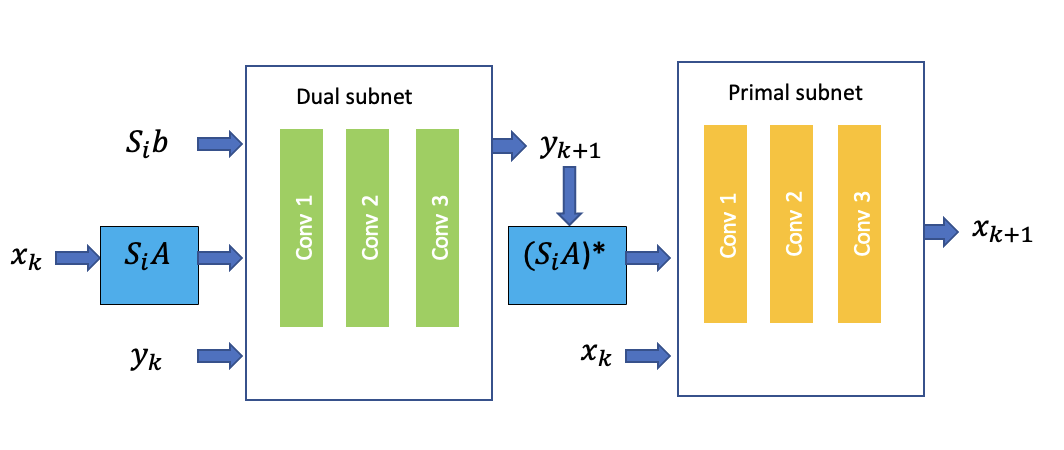}}
    \caption{One simple example of the practical choices for the building blocks of one layer of our LSPD network. Both dual and primal subnetworks are consist of 3 convolutional layers. The dual subnet has 3 input channels concatenating $[S_ib, S_iAx_k, y_k]$, while the primal subnet has 2 input channels for $[(S_iA)^Ty_{k+1}, x_k]$}

\end{figure} 

Note that the LSPD network we presented above is corresponding to a variant of SPDHG without variance-reduction, since in each layer it considers only a stochastic gradient of a block of dual variable, while the SPDHG of \cite{chambolle2018stochastic} has a variance-reduction mechanism, by averaging the stochastic gradient of the dual variables of earlier steps leading to accelerated convergence. Hence we also propose a variant of the LSPD by fully utilizing this variance-reduction technique:
 \begin{eqnarray*}
 && \mathrm{\textbf{Learned Stochastic Primal-Dual}} \\&&\mathrm{\textbf{ with Variance-Reduction (LSPD-VR)}} \\&&- \mathrm{Initialize}\ x_0 \in \mb{R}^d, \ y_0 \in \mb{R}^{p/m},\ h_1,...,h_m =\textbf{0}_{p/m}\\
 &&\mathrm{For} \ \ \ k = 0, 1, 2,...,  K-1\\
&&\left\lfloor
\begin{array}{l}
i = \mod(k,m);\\ \mathrm{(or\ pick\ i\ from\ [0, m-1]\ uniformly\ at\ random)}\\
y_{k+1} = \D_{\te_d^k} (y_k, \sigma_k, (S_iA) {x}_k , S_ib);\\
h_i = (S_iA)^T y_{k+1};\\
x_{k+1} = \Pm_{\te_p^k}(x_k, \tau_k, \sum_{j=0}^{m-1} h_j);\\
\end{array}
\right.
 \end{eqnarray*}
The LSPD-VR can be view as an extenstion of LSPD with dense skip-connections across layers. From an optimization view-point, we know that non-variance-reduced stochastic gradient methods typically converge faster than variance-reduced ones in early iterations. However, if the number of iteration is large, then variance-reduced methods would converge much faster in terms of optimization accuracy. Hence we suspect that when $K$ is large, LSPD-VR would be a better choice, while the LSPD should be more advantageous when $K$ is relatively small.

\section{Theoretical Analysis of LSPD Network}

In this section we provide theoretical recovery analysis of a subclass of the LSPD framework presented in the previous section. From this motivational analysis, we aim to demonstrate the reconstruction guarantee of a dual-free, recurrent version of the LSPD, and compare it with the recovery guarantee of the LPD derived under the same setting.
 \begin{eqnarray*}
 && \mathrm{\textbf{Simplified LSPD}} - \mathrm{Initialize}\ x_0 \in \mb{R}^d \ y_0 \in \mb{R}^{p/m}\\
 &&\mathrm{For} \ \ \ k = 0, 1, 2,...,  K-1\\
&&\left\lfloor
\begin{array}{l}
\mathrm{Pick\ i\ from\ [0, m-1]\ uniformly\ at\ random};\\
y_{k+1} = S_iA {x}_k  - S_ib;\\
x_{k+1} = \Pm_{\te_p}(x_k - \tau (S_iA)^T y_{k+1});\\
\end{array}
\right.
 \end{eqnarray*}
We analyze here a simplified variant of LSPD, where we set the dual sub-networks\footnote{This option of dual subnetwork has been also considered in the original LPD paper's numerics.} to be a simple subtraction operation between the dual update $S_iA {x}_k$ and the measurement $S_ib$, and meawhile we set the primal subnetworks to have the same weights, trained separately as an approximate projection operator towards some image manifold $\mathcal{M}$. Hence, this simplified version can be view as a learned proximal SGD. A typical example for this type of construction of unrolling network can be found in \citep{rick2017one}.

Notably, we choose a trivial parameterization for the dual sub-network to make the error analysis feasible. Our ongoing work includes the analysis and understanding of learning the dual subnetwork which is effectively the learning of an adaptive $f_b$ loss.

\subsection{Generic Assumptions}
We list here the assumptions we make in our motivational analysis of the simplified LSPD.
\begin{myh}
\textbf{(Approximate projection)} We assume that the primal subnetwork of the Simplified LSPD is a $\ve$-approximate projection towards a manifold $\M$: 
\begin{equation}
    \Pm_{\te_p}(x) = e(x) + P_\M(x),
\end{equation}
where:
\begin{equation}
    P_\M(x) := \arg \min_{z \in \M} \|x - z\|_2^2,
\end{equation}
and,
\begin{equation}
    \|e(x)\|_2 \leq \ve , \ \ \forall x \in \mb{R}^d.
\end{equation}
\end{myh}
Here we model the primal subnetwork to be a $\ve$-projection towards a manifold. Note that in practice the image manifold $\M$ typically form a non-convex subset. We also make conditions on the image manifold as the following:
\begin{myh}
\textbf{(Interpolation)}We assume the ground-truth image $x^\dagger \in \M$, where $\M$ is a closed set.
\end{myh}
With this condition on the manifold, we further assume restricted eigenvalue (restricted strong-convexity) condition which is necessary for robust recovery.
\begin{myh}
\textbf{(Restricted Eigenvalue Condition)} We define a descent cone $\mathcal{C}$ at point $x^\dagger$ as:
 \begin{equation}
    \mathcal{C} := \left\{v \in \mathbb{R}^d |\  v = a(x - x^\dagger) , \forall a \geq 0, x \in \mathcal{M} \right\},
\end{equation}
and the restricted strong-convexity constant $\mathcal{\mu}_c$ to be the largest positive constant satisfies the following: 
 \begin{equation}
    \frac{1}{n} \|Av\|_2^2 \geq \mu_c \|v\|_2^2 ,\ \ \forall v \in \mathcal{C}.
 \end{equation}
 and the restricted smoothness constant $L_c$ to be the smallest positive constant satisfies:
 \begin{equation}
    \frac{1}{q} \|S_iAv\|_2^2 \leq L_c \|v\|_2^2 ,\ \ \forall v \in \mathcal{C}. \ \forall i \in [m]
 \end{equation} 
\end{myh}
The restricted eigenvalue condition is standard and crucial for robust estimation guarantee for linear inverse problems, i.e., for a linear inverse problem to be non-degenerate, such type of condition must hold \citep{oymak2017sharp, agarwal2012fast, 2012_Chandrasekaran_Convex}. For example, in sparse-recovery setting, when the measurement operator is a Gaussian map (compressed-sening measurements) and $x^\dagger$ is $s$-sparse, one can show that $\mu_c$ can be as large as $O(1 - \frac{s \log d}{n})$ \citep{oymak2017sharp}. In our setting we would expect an even better $\mu_c$, since the mainifold of certain classes of real-world images should have much smaller covering numbers compare to the sparse set.

\subsection{Estimation error bounds of Simplified LSPD}
With the assumptions presented in the previous subsetion, here we provide the recovery guarantee of a $K$ layer simplified LSPD network on linear inverse problem where we have $b = Ax^\dagger$. Denoting $L_s$ to be the smallest constant satisfying:
\begin{equation}
    \frac{1}{q}\|S_iAv\|_2^2 \leq L_s \|v\|_2^2, \ \ \forall v \in \mb{R}^d, i \in [m],
\end{equation}
we can have the following result:
\begin{theorem}\label{thm1}
(Upper bound) Assuming \textbf{A.1-3}, let $\tau = \frac{1}{qL_s}$ and $b = Ax^\dagger + w$, the output of a $K$ layer Simplified LSPD network has the following guarantee for the estimation of $x^\dagger$:
\begin{equation}
    \E\|x_K - x^\dagger\|_2 \leq \alpha^{K} \|x_0 - x^\dagger\|_2+ \frac{1 - \alpha^{{K}}}{1 - \alpha} (\ve + \delta),
\end{equation}
where $\alpha = 2(1 - \frac{\mu_c}{L_s})$, and\footnote{We denote $\mc{B}^d$ and $\mb{S}^{d-1}$ as the unit ball and unit sphere in $\mb{R}^d$ around origin, respectively.},
\begin{equation}
    \delta:= 2\tau\E\sup_{v \in \mathcal{C} \cap \mathcal{B}^{d}, i \in [m]} v^TA^TS_i^TS_iw
\end{equation}
\end{theorem}
\footnote{For the noiseless case $b = Ax^\dagger$, $\delta = 0$.}When the restricted eigenvalue $\mu_c$ is large enough such that $\alpha < 1$, the simplified LSPD has a linear convergence in estimation error, up to $\frac{\ve}{1 - \alpha}$ only depending the approximation accuracy of the primal-subnet in terms of projection. For many inverse problems for example CT/PET tomographic imaging we have $L_s \approx L_f$ where $L_f$ being the largest eigenvalue of $\frac{1}{n}A^TA$, and in these tasks the same convergence rate in Theorem \ref{thm1} apply for both LSPD and LPD. This suggests the tremendous potential of computational saving of LSPD over full batch LPD.

One the other hand, using similar technique we can provide a complementing lower bound for the estimation error of LSPD:
\begin{theorem}
 (Lower bound.) Under the same conditions of Theorem \ref{thm1}, if we further assume the constraint set $\M$ is convex, for any $\gamma > 0$, $\exists R(\gamma)$, if $\|x_0 - x^\dagger\|_2 \leq R(\gamma)$, the estimation error of the output of LSPD satisfies the lower bound:
 \begin{equation}
     \E\|x_K - x^\dagger\|_2 \geq (1 - \gamma)^K (1 - \frac{L_c}{L_s})^K \|x_0 - x^\dagger\|_2 - \frac{L_s}{L_c} \ve
 \end{equation}
\end{theorem}
To have a clearer view of the upper and lower bounds, we make a case study for quantifying the bounds in the next subsection.
\subsection{Case study: generalized Gaussian map}
To make a quantification of the bounds, we consider the following special case of foward operator:
\begin{equation}\label{cgau}
    A = GB, \forall i \in [n], j \in [d], G_{ij} \sim \mathcal{N}(0, \frac{1}{n}).
\end{equation}
The choice $B = I$ makes $A$ a Gaussian map. Note that for any overdetermined $A$, such a construction always exists. Now we introduce the notion of Gaussian width \cite{2012_Chandrasekaran_Convex} measuring the size of the constraint set $\M$:
\begin{definition}
(Gaussian Width) The Gaussian width of a set $\M$ is defined as:
\begin{equation}
    \W(\M) := \E_u \left( \sup_{v \in \M} \langle v, u \rangle \right)
\end{equation}
\end{definition}
We start by presenting a simple extension of the Gordon's escape-through-a-mesh Lemma \citep[Lemma 21]{oymak2017sharp} adapted to our setting: 
\begin{lemma}
(Generalized Escape-through-a-mesh) For the construction $A = GB$, denote the largest singular value and smallest singular value of $B$ as $\sigma_a$ and $\sigma_b$, respectively, denoting $u_n \sim \mc{N}(0, I_n)$, $p_n = \E\|u_n\|_2 \approx \frac{\sqrt{2}\Gamma(\frac{n+1}{2})}{\Gamma(\frac{n}{2})} \approx \sqrt{n}$, $\forall v \in B\mc{C}\cap \mb{S}^{n-1}$, with probability at least $1 - e^{-\frac{\theta^2}{2}}$, we have the following bound:
\begin{equation}
    \frac{\|Av\|_2}{\sigma_b \|v\|_2} \geq p_n - \W(B\mc{C} \cap \mb{S}^{n-1}) - \theta.
\end{equation}
Meanwhile, with probability at least $1 - m e^{-\frac{\theta^2}{2}}$, we have the following bound:
\begin{equation}
    \frac{\|S_iAv\|_2}{\sigma_a\|v\|_2} \leq p_q +  \W(B\mc{C} \cap \mb{S}^{n-1}) + \theta, \forall i \in [m].
\end{equation}
\end{lemma}
With this adapted Gordon's Lemma, we can quantify the bounds in the previous section with exponentially high probability.
\begin{theorem}
Assuming \textbf{A.1-3}, let $\tau = \frac{1}{qL_s}$ and $b = Ax^\dagger$, where $A = GB$ as in (\ref{cgau}), the output of a $K$-th layer Simplified LSPD network has the following guarantee for the estimation of $x^\dagger$ with probability at least $1-mKe^{-\frac{\theta^2}{2}}$:
\begin{equation}
    \E\|x_K - x^\dagger\|_2 \leq \alpha_U^{K} \|x_0 - x^\dagger\|_2+ \frac{1}{1 - \alpha_U} \ve,
\end{equation}
where:
\begin{equation}
    \alpha_U := \kappa_\M \left( 1 - \frac{\sigma_b q(p_n - \W(B\mc{C} \cap \mb{S}^{n-1}) - \theta)^2}{\sigma_a n(p_q + \sqrt{d} + \theta)^2} \right) 
\end{equation}
where we denote $\kappa_\M = 2$ if $\M$ is non-convex, $\kappa_\M = 1$ if otherwise. If $\M$ is a convex set, for any $\gamma > 0$, $\exists R(\gamma)$, if $\|x_0 - x^\dagger\|_2 \leq R(\gamma)$, we can have the lower bound with probability at least $1 - me^{-\frac{\theta^2}{2}}$:
\begin{equation}
    \E\|x_K - x^\dagger\|_2 \geq (1-\gamma)^K\alpha_L^{K} \|x_0 - x^\dagger\|_2 - \frac{1}{1 - \alpha_L} \ve,
\end{equation}
where:
\begin{equation}
    \alpha_L :=  1 - \frac{\sigma_b (p_q + \W(B\mc{C} \cap \mb{S}^{n-1}) + \theta)^2}{\sigma_a (p_q - \sqrt{d} - \theta)^2} 
\end{equation}
\end{theorem}

\textbf{Remark.}For non-convex $\M$, a non-vacuous upper bound with $\alpha_U < 1$ can only be obtainable in a regime where $\sigma_a \approx \sigma_b$ (which means the operator $A$ is close to a Gaussian map up to nearly orthogonal transformation), and where we have a large number of measurement data ($n \gg d$). Noting that $\W(B\mc{C} \cap \mb{S}^{n-1}) \leq \sqrt{d}$, the convergence rate $\alpha_U$ in this data-rich regime can be approximated as:
\begin{equation}
\begin{aligned}
       \alpha_U &\approx \kappa_\M \left( 1 - \frac{(\sqrt{n} - \W(B\mc{C} \cap \mb{S}^{n-1}))^2}{(\sqrt{n} + \sqrt{\frac{nd}{q}})^2} \right)\\
       &\approx \kappa_\M \left( 1 - \frac{n}{(\sqrt{n} + \sqrt{\frac{nd}{q}})^2} \right),
\end{aligned}
\end{equation}
and we observe that as long as the minibatch size $q > d/(\sqrt{2} - 1) + O(1)$ we are ensured to have $\alpha_U < 1$ with exponentially high probability. Meanwhile the lower bound confirms the convergence rate $\alpha_U$ is indeed sharp (worst-case optimal) up to a constant. If we choose a large minibatch\footnote{In our numerical experiments we indeed choose a large minibatch size $q = \frac{n}{4}$, which means we make 4 subsets from the forward operator $A$.} $q = \frac{n}{O(1)}$ in this data-rich regime ($n \gg d$), we are ensured to have the same convergence rate of the full batch LPD network under the same parameterization. Hence our motivational analysis suggests that a \say{free-lunch} of computational reduction using our LSPD framework is obtainable. From the analysis above, we also obtain insights that the minibatch size should be relatively large in order to maintain good performance.

\begin{figure*}[t]
   \begin{center}
    {\includegraphics[trim=80 80 15
    55,clip,width=0.9\textwidth]{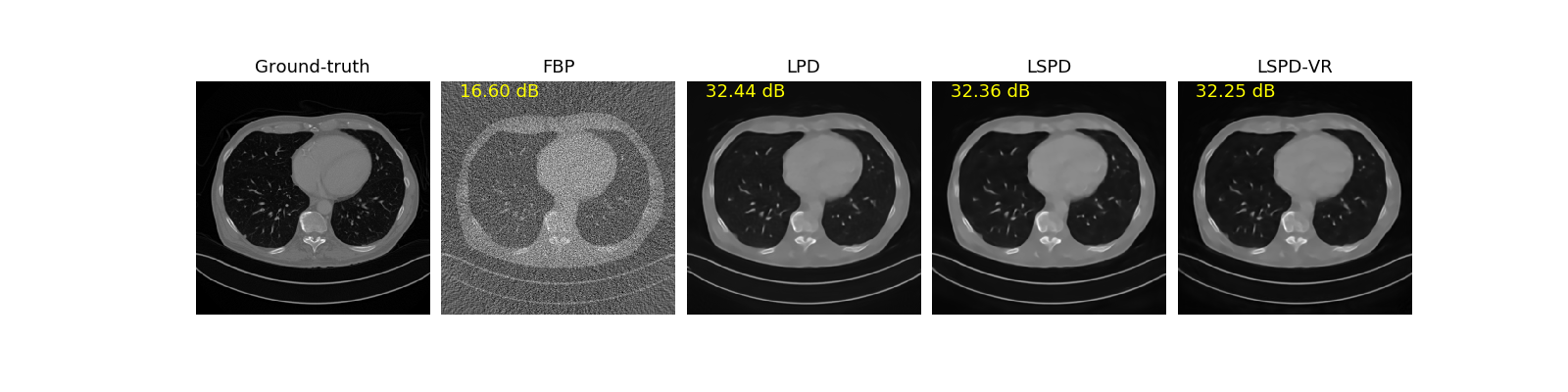}}
    {\includegraphics[trim=80 130 15 135,clip,width=0.9\textwidth]{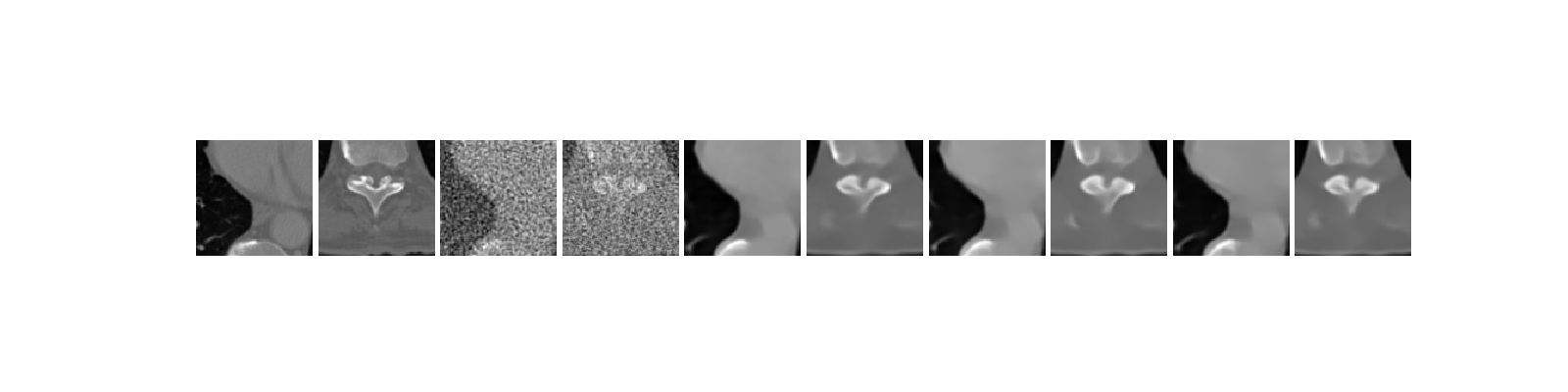}}
        {\includegraphics[trim=80 80 15 55,clip,width=0.9\textwidth]{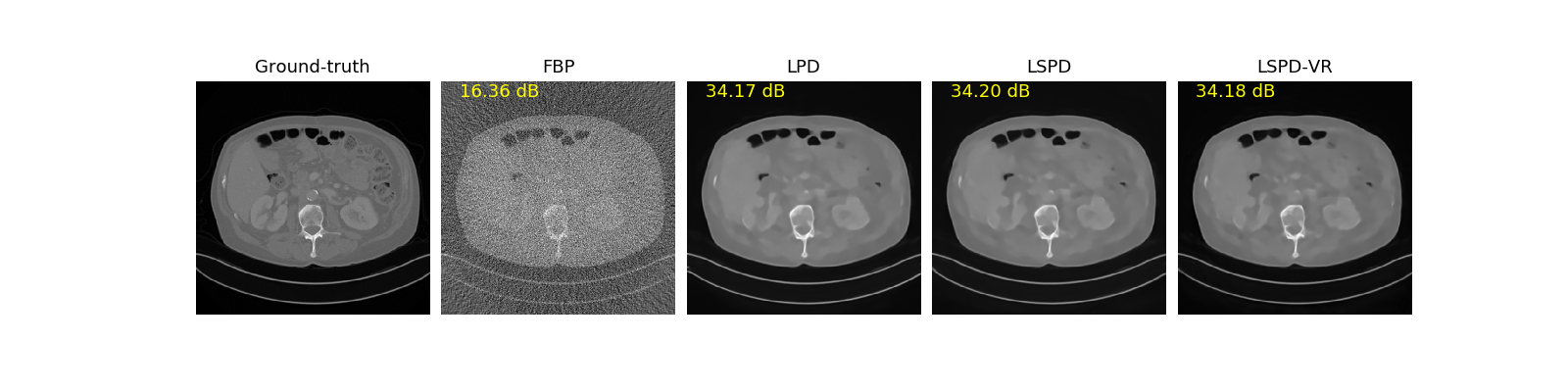}}
            {\includegraphics[trim=80 130 15 135,clip,width=0.9\textwidth]{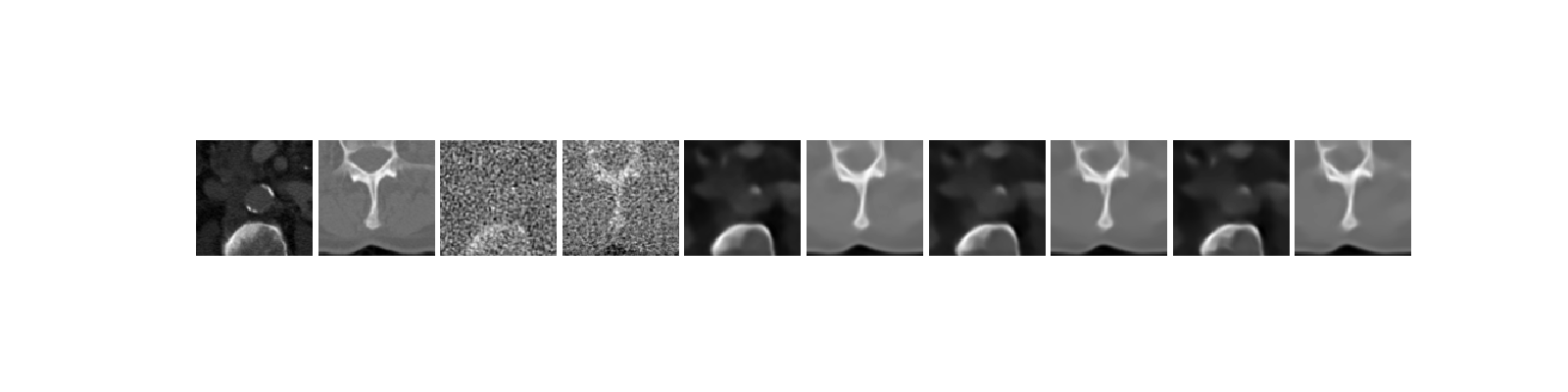}}
    \end{center}
   \caption{Examples for Low-dose CT on the test set of Mayo dataset. We can observe that our LSPD networks achieve the same reconstruction performance as the full-batch LPD}
    \label{f3}
\end{figure*}

\section{Training of LSPD}
\subsection{Supervised end-to-end training}
The most basic training approach for LSPD is the end-to-end supervised training where we consider fully paired training samples of measurement and the \say{ground-truth} -- which is typically obtained via a reconstruction from high-accuracy and abundant measurements. We take the initialization of LSPD as a \say{filtered back-projection} $x^0 = A^\dagger b$. Let $\te$ be the set of parameters $\te := \{\te_p^k, \te_d^k, \tau_k, \sigma_k\}_{k=0}^{K-1}$, applying the LSPD network on some measurement $b$ can be written as $\F_\te(b)$, the training objective can typically be written as:
\begin{equation}\label{sup_loss}
    \te^\star \in \arg\min_{\te} \sum_{i=1}^N \|x_i^\dagger - \F_{\te}(b_i, x^0_i)\|_2^2,
\end{equation}
where we denote by $N$ the number of paired training examples. Since the paired high-quality ground-truth data is expensive and idealistic in practice, we would prefer more advanced methods for training which have relaxed requirements on available training data. In Appendix, we present unsupervised training results for the unrolling networks using only measurement data.

\subsection{Fine-tuning step by instance-adaptation for out-of-distribution reconstruction}

In medical imaging practice, we may encounter the scenario where the input data come from slightly different measurement modality or measurement noise distribution than the training data we used for training the unrolling network. In such a setting, instead of retraining the whole network, it is more desirable to fine-tune the network to adapt itself to this input data \citep{yu2021empirical,adler2021task}. This process is known as instance-adaptation \citep{vaksman2020lidia,tachella2020neural} which was applied in boosting denoising network. Let $g \in \G$ where $\G$ is a group of transformations (for example, rotation transformation), and we assume that the operation of forward pass of the network $\F_\theta(A\    \cdot)$ is equivariant for this group of transformation:
\begin{equation}
    T_g\F_\theta(Ax) = \F_\theta(AT_g(x)), \ \ \forall x \in X \subset \R^d,
\end{equation}
where $T_g$ is a unitary matrix such that $T_gx \in X, \ \forall x \in X \subset \R^d$. Such a condition is true for CT/PET if we consider rotation operation \citep{celledoni2021equivariant}. Our post-processing method can be describe as the following: taking a pre-trained LSPD network $\F_{\te^\star}(\cdot)$ apply it on the input data $b_{\mr{in}}$, and run a few steps of Adam optimizer on the following self-supervised objective (initialized with $\te^\star$):
\begin{equation}\label{ia_ei_obj}
\begin{aligned}
   \te_a^\star \approx \arg\min_{\te}& \E_{g} \{\|b_{\mr{in}} - A(\F_\te(b_{\mr{in}}))\|_2^2 \\&+ \lambda \|T_g\F_\te(b_{\mr{in}}) - \F_\te(AT_g\F_\te(b_{\mr{in}}))\|_2^2 \},
\end{aligned}
\end{equation}
where we adapted the equivariant regularization term proposed in \citep{chen2021equivariant} to our setting for data-augmentation, and then reconstruct the underlying image as $x_a^\star = \F_{\te_a^\star}(b_{\mr{in}})$. Note that, this post-processing is not very recommended for classic full-batch unrolling network, since it will require a number of calls on full forward and adjoint operator, which is computationally inefficient. However, for our LSPD network, such a limitation can be significantly mitigated since we only use subsets of these operators. We present some numerical results of this optional step on sparse-view CT in the Appendix.

\begin{table*}[t]\label{ld_table}
\caption{Low-dose CT testing results for LPD, LSPD and LSPD-VR networks on Mayo dataset, with supervised training}
\label{sample-table}
\vskip 0.15in
\begin{center}
\begin{small}
\begin{sc}
\begin{tabular}{lcccr}
\hline
Method & $\#$ calls on $A$ and $A^T$& PSNR & SSIM  \\
\hline

&&&\\
Filtered Backprojection (FBP) & - & 16.4148& 0.0929
\\
FBPConvNet & - & 30.8942& 0.8508
\\
LPD \textit{(12 layers)} & 24& 34.0913 & 0.8920  \\

LSPD  \textit{(12 layers)} & 6& 34.0469 & 0.8845\\

LSPD-VR \textit{(12 layers)} & 6 & 33.9801 & 0.8793\\

\hline
\end{tabular}
\end{sc}
\end{small}
\end{center}
\vskip -0.1in
\end{table*}

\section{Numerical Results}


In this subsection we present numerical results of our proposed networks for low-dose X-ray CT. In real world clinical practice, the low dosage CT is widely used and highly recommended, since the intense exposures to the X-ray could significantly increase the risk of inducing cancers \citep{mason2018quantitative}. The low-dose CT takes a large number of low-energy X-ray views, leading to huge volumes of noisy measurements. This makes the reconstruction schemes struggle to achieve efficient and accurate estimations. In our X-ray CT experiments we use the standard Mayo-Clinic dataset \citep{mccollough2016tu} which contains 10 patients' 3D CT scans. We use 2111 slices (from 9 patients) of 2D images sized $512 \times 512$ for training and 118 slices of the remaining 1 patient for testing. We use the ODL toolbox \citep{adler2018learned} to simulate fan beam projection data with 800 equally-spaced angles of views (each view includes 400 rays). The fan-beam CT measurement is corrupted with Poisson noise: $b \sim \mr{Poisson}(I_0 e^{-Ax^\dagger})$, where we make a low-dose choice of $I_0=3.5 \times 10^4$.  This formula is used to simulate the noisy projection data according to the Beer-Lambert law, and to linearize the measurements, we consider the log data.

In our LSPD and LSPD-VR networks we block-partition (according to the angles) the forward/adjoint operators and data into 4 subsets. Our networks has 12 layers\footnote{each layer of LSPD includes a primal and a dual subnetwork with 3 convolutional layers with kernel size $5 \times 5$ and 32 channels, same for LPD.} hence correspond to 3 data-passes, which means it takes only 3 calls in total on the forward and adjoint operator. We compare it with the learned primal-dual (LPD) which has 12 layers, corresponding to 12 calls on the forward and adjoint operator. We train all the networks with 50 epochs of Adam optimizer \citep{kingma2014adam} with batch size 1, in the supervised manner.

\begin{figure}[t]
   \begin{center}
    {\includegraphics[trim=20 20 15 15,clip,width=0.45\textwidth]{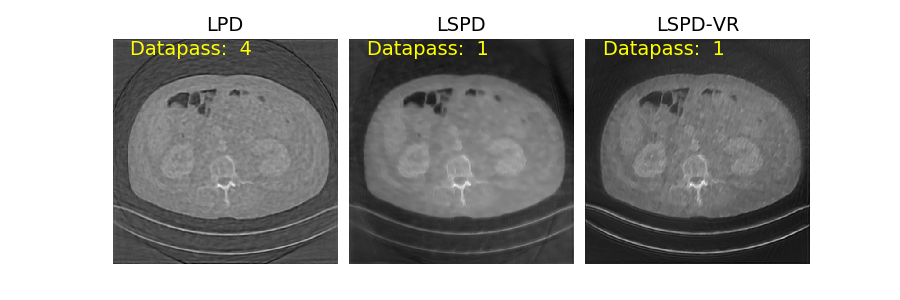}}
    {\includegraphics[trim=20 20 15 15,clip,width=0.45\textwidth]{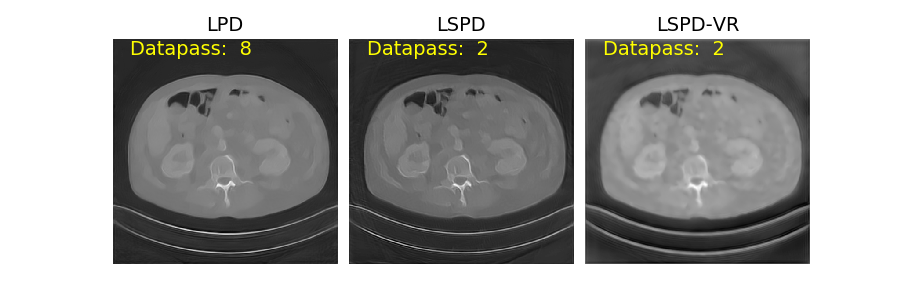}}
        {\includegraphics[trim=20 20 15
    15,clip,width=0.45\textwidth]{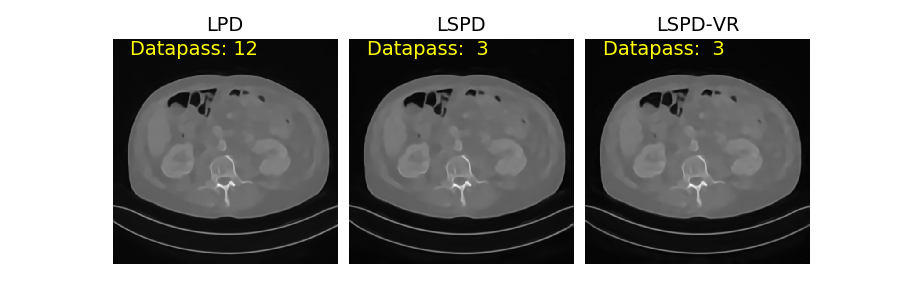}}
    \end{center}
   \caption{Example for intermediate layer outputs for Low-dose CT on the test set of Mayo dataset. We can observe that LSPD/LSPD-VR achieves competitive reconstruction quality with LPD across intermediate layers.}
    \label{f33}
\end{figure}

We present the performance of the LPD. LSPD, and LSPD-VR on the test set in Table 1, and some illustrative examples in Figure 2 for a visual comparison. We also present the results of the classical Filtered-Backprojection (FBP) algorithm, which is widely used in clinical practice. We can observe from the FBP baseline, due to the challenging extreme low-dose setting, the FBP reconstruction fails completely. This can be partially addressed by U-Net postpocessing (FBPConvNet, \citet{jin2017deep}), whose parameter size is one order of magnitute larger than our unrolling networks. Next we turn to the learned reconstruction results. From the numerical results, we found out that our LSPD and LSPD-VR networks both achieve almost the same reconstruction accuracy compare to LPD baseline in terms of PSNR (peak signal-to-noise ratio) and SSIM (structural similarity index, \citep{wang2004image}) measures, with requiring only a fraction of the computation of the forward and adjoint operators.

 Moreover we found that the LSPD-VR does not seem to improve upon the LSPD, which suggests that the variance-reduction tricks which made huge progress in the field of stochastic optimization, may not be effective in designing the stochastic deep unrolling networks. The main possible reason here should be the fact that we use large minibatches of the operators, which makes the variance of gradient estimate reasonably small. We conjecture that if small minibatches are used, the variance-reduction trick could be effective -- we leave this study as a future work.

In our Appendix, we present additional results on another widely applied modality in clinical practice -- the sparse-view CT, where we take much fewer amount of normal-dose measurements. Different to the low-dose CT, the main challenge of sparse-view CT is the ill-poseness of the inverse problems, that the measurement operator is highly under-determined with a non-trivial null-space. Meanwhile we also present in the Appendix our numerical results on instance-adaptation in out-of-distribution reconstruction for sparse-view CT, demonstrating again the effectiveness of our approach.

\bibliography{main.bib}
\bibliographystyle{icml2022}

\newpage
\appendix
\onecolumn

\section{Proof of Theorem 3.1}

In this proof we utilize several projection identities from \citep{oymak2017sharp}. We list them here first for completeness. The first one would be the cone-projection identity:
\begin{equation}
    \|P_\mathcal{C}(x)\|_2 = \sup_{v \in \mathcal{C} \cap \mathcal{B}^d} v^Tx,
\end{equation}
where $\mathcal{B}^d$ denotes the uni-ball in $\mb{R}^d$. The second one is the shift projection identity regarding that the Euclidean distance is preserved under translation:
\begin{equation}
    P_\M(x + v) - x = P_{\M - x} (v).
\end{equation}
Now if $0  \in \M - x$, we can have the third projection identity which is an important result from geometric functional analysis \citep[Lemma 18]{oymak2017sharp}:
\begin{equation}
    \|P_\D(x)\|_2 \leq \kappa_\D \|P_\mc{C}(x)\|_2,
\end{equation}
where:
\begin{equation}
    \kappa_\D = \left\{ \begin{array}{ll}
        1 &  \mbox{if $\mathcal{D}$ is convex}\\
       2 &  \mbox{if $\mathcal{D}$ is non-convex}
        \end{array}
        \right.
    \end{equation}
where $\D$ is a potentially non-convex closed set included by cone $\mc{C}$. On the other hand, utilizing a simplified result of \citep{pmlr-v97-qian19b} with partition minibatches, we have:
\begin{equation}
    \mathbb{E}_S(\|A^TS^TSA(x - z)\|_2^2) \leq 2 L_s (\frac{q^2}{2n}\|Ax - b\|_2^2 - \frac{q^2}{2n}\|Az- b\|_2^2 - q^2 \langle \triangledown f(z), x-z \rangle).
\end{equation}
Then for the case of noisy measurements $b= Ax^\dagger + w$, following similar procedure we can have:
\begin{eqnarray}\label{esmooth}
&&\mathbb{E}_S(\|A^TS^TSA(x - z)\|_2^2)\\ && \leq 2 L_s (\frac{q^2}{2n}\|Ax - b\|_2^2 - \frac{q^2}{2n}\|Az- b\|_2^2 - q^2 \langle  \frac{1}{n} A^T(Az - b), x-z \rangle)\\
&& \leq \frac{q^2L_s}{n}(\frac{1}{2}\|A(x - x^\dagger) - w\|_2^2 - \|w\|_2^2 + \langle w, A(x- x^\dagger)\rangle)\\
&&=\frac{q^2L_s}{n}\|A(x-x^\dagger)\|_2^2
\end{eqnarray}
For $k$-th layer of simplified LSPD we have the following:
\begin{eqnarray*}
&&\|x_{k+1}-x^\dagger\|_2\\
&=& \|\mathcal{P}_{\theta_p}(x_k- \tau A^T{S_i}^Ty_k)-x^\dagger\|_2\\
&\leq& \|P_\M(x_k- \tau A^T{S_i}^Ty_k)-x^\dagger\|_2 + \|e(x_k- \tau A^T{S_i}^Ty_k)\|_2\\
&=& \|P_{\M-x^\dagger}(x_k-x^\dagger- \tau A^T{S_i}^Ty_k)\|_2 + \|e(\Bar{x_k})\|_2,
\end{eqnarray*}
where we used the second cone-projection identity and denoted $\Bar{x_k}:= x_k- \tau A^T{S_i}^Ty_k$. Then by applying the first and third identity we can continue:
\begin{eqnarray*}
&&\|x_{k+1}-x^\dagger\|_2\\ 
&\leq& 2\|P_{\mathcal{C}}(x_k-x^\dagger - \tau A^T{S_i}^Ty_k)\|_2 + \|e(\Bar{x_k})\|_2\\ 
&=& 2\sup_{v \in \mathcal{C} \cap \mathcal{B}^{d}} v^T(x_k-x^\dagger - \tau A^T{S_i}^Ty_k) + \|e(\Bar{x_k})\|_2\\
&\leq& 2\sup_{v \in \mathcal{C} \cap \mathcal{B}^{d}} v^T(x_k-x^\dagger - \tau A^TS_i^T(S_iAx_k-S_ib)) + \|e(\Bar{x_k})\|_2\\
&=& 2\sup_{v \in \mathcal{C} \cap \mathcal{B}^{d}} v^T[x_k-x^\dagger - \tau A^TS_i^T(S_iAx_k-S_i(Ax^\dagger + w))] + \|e(\Bar{x_k})\|_2\\
&\leq& 2\sup_{v \in \mathcal{C} \cap \mathcal{B}^{d}} v^T[(I - \tau A^T{S_i}^T{S_i}A)(x_k-x^\dagger)] + 2\tau\sup_{v \in \mathcal{C} \cap \mathcal{B}^{d}} v^TA^TS_i^TS_iw + \|e(\Bar{x_k})\|_2\\
&\leq& 2\|(I - \tau A^T{S_i}^T{S_i}A)(x_k-x^\dagger)\|_2+ 2\tau\sup_{v \in \mathcal{C} \cap \mathcal{B}^{d}} v^TA^TS_i^TS_iw + \|e(\Bar{x_k})\|_2.
\end{eqnarray*}

Denote $\Bar{x_k}:= x_k- \tau A^T{S_i}^Ty_k$, and take expectation, then we have:
\begin{eqnarray*}
&&\E(\|x_{k+1}-x^\dagger\|_2)\\
&\leq& 2\E(\|(I - \tau A^T{S_i}^T{S_i}A)(x_k-x^\dagger)\|_2) + \|e(\Bar{x_k})\|_2 + + 2\tau\E\sup_{v \in \mathcal{C} \cap \mathcal{B}^{d}} v^TA^TS_i^TS_iw\\
&\leq& 2\sqrt{ \E(\|(I - \tau A^T{S_i}^T{S_i}A)(x_k-x^\dagger)\|_2^2)} + \|e(\Bar{x_k})\|_2 + 2\tau\E\sup_{v \in \mathcal{C} \cap \mathcal{B}^{d}} v^TA^TS_i^TS_iw\\
&=& 2\sqrt{ \E( \|x_k-x^\dagger\|_2^2-2\tau\|{S_i}A(x_k-x^\dagger)\|_2^2+\tau^2\|A^T{S_i}^T{S_i}A(x_k-x^\dagger)\|_2^2)} + \|e(\Bar{x_k})\|_2 + 2\tau\E\sup_{v \in \mathcal{C} \cap \mathcal{B}^{d}} v^TA^TS_i^TS_iw\\
\end{eqnarray*}
Now denoting:
\begin{equation}
    \delta:= 2\tau\E\sup_{v \in \mathcal{C} \cap \mathcal{B}^{d}, i \in [m]} v^TA^TS_i^TS_iw
\end{equation}
since (\ref{esmooth}), we can continue:
\begin{eqnarray*}
&&\E(\|x_{k+1}-x^\dagger\|_2)\\
&\leq& 2\sqrt{  \|x_k-x^\dagger\|_2^2-2\frac{\tau q}{n}\|A(x_k-x^\dagger)\|_2^2+\frac{\tau^2q^2L_s}{n} \|A(x_k - x^\dagger)\|_2^2 } + \|e(\Bar{x_k})\|_2 + \delta\\
&\leq& 2\sqrt{  \|x_k-x^\dagger\|_2^2-(2\tau q-2L_s\tau^2 q^2)\cdot \frac{1}{n} \|A(x_k - x^\dagger)\|_2^2} + \|e(\Bar{x_k})\|_2+ \delta
\end{eqnarray*}
and then due to Assumption A.3 the Restricted Eigenvalue Condition we have:
\begin{eqnarray*}
\E(\|x_{k+1}-x^\dagger\|_2)
&\leq& 2\sqrt{  \|x_k-x^\dagger\|_2^2-(2\tau q\mu_c - 2L_s \mu_c\tau^2 q^2 ) \|x_k - x^\dagger\|_2^2}  + \|e(\Bar{x_k})\|_2+ \delta\\
&=&2\sqrt{1- 2\mu_c\tau q+ 2L_s \mu_c\tau^2q^2}\|x_k-x^\dagger\|_2  + \|e(\Bar{x_k})\|_2+ \delta\\
&\leq& 2(1-\frac{\mu_c}{L_s})\|x_k-x^\dagger\|_2  + \|e(\Bar{x_k})\|_2 + \delta
\end{eqnarray*}
Then let $\alpha=2(1-\frac{\mu_c}{L_s})$, and noting that we have assumed $\|e(x)\|_2 \leq \ve$, by the tower rule we get:
 \begin{eqnarray*}
     \E(\|x_k-x^\dagger\|_2) &\leq& \alpha^{{k}}\|x_0-x^\dagger\|_2 + \frac{(1- \alpha^{k})}{1-\alpha} (\ve + \delta).
 \end{eqnarray*}
Thus finishes the proof.

\section{Proof for Theorem 3.2}

For proving the lower bound we will need to assume the constraint set $\M$ to be convex and apply a know result provide in \citet[Lemma F.1]{oymakbabak2016sharp}, that for a closed convex set $\D:= \M - x^\dagger$ containing the origin, given any $a, \gamma \in (0,1]$ there exist a positive constant $C$ such that for any $v$ satisfies $\|\mc{P}_\mc{C}(v)\|_2 \geq a \|v\|_2$ and $\|v\|_2 \leq c$, we can have:
\begin{equation}\label{lemo}
    \frac{\|P_\D(v)\|_2}{\|P_\mc{C}(v)\|_2} \geq 1 - \gamma.
\end{equation}
Since in A.2 we assume the ground truth $x^\dagger \in \M$  we know that $0 \in \M - x^\dagger$ hence the above claim is applicable.
For $k$-th layer of simplified LSPD we have the following:
\begin{eqnarray*}
&&\|x_{k+1}-x^\dagger\|_2\\
&=& \|\mathcal{P}_{\theta_p}(x_k- \tau A^T{S_i}^Ty_k)-x^\dagger\|_2\\
&\geq& \|P_\M(x_k- \tau A^T{S_i}^Ty_k)-x^\dagger\|_2  - \|e(x_k- \tau A^T{S_i}^Ty_k)\|_2\\
&=& \|P_{\M-x^\dagger}(x_k-x^\dagger- \tau A^T{S_i}^Ty_k)\|_2 - \|e(x_k- \tau A^T{S_i}^Ty_k)\|_2.
\end{eqnarray*}
Now due to (\ref{lemo}) we can continue:
\begin{eqnarray*}
\|x_{k+1}-x^\dagger\|_2 &\geq& (1 - \gamma)\|P_{\mc{C}}(x_k-x^\dagger- \tau A^T{S_i}^Ty_k)\|_2 - \varepsilon\\
&=& (1 - \gamma)\|P_{\mc{C}}[(I- \tau A^T{S_i}^TS_iA)(x_k-x^\dagger)]\|_2 - \varepsilon\\
&=& (1 - \gamma)\sup_{v \in \mc{C} \cap \mb{S}^{d-1}} v^T(I- \tau A^T{S_i}^TS_iA)(x_k-x^\dagger)- \varepsilon\\
&\geq& (1 - \gamma) \frac{x_k - x^\dagger}{\|x_k - x^\dagger\|_2}(I- \tau A^T{S_i}^TS_iA)(x_k-x^\dagger)- \varepsilon\\
&=& (1 - \gamma) \|x_k - x^\dagger\|_2\left(1 - \tau\frac{\|S_iA(x_k - x^\dagger)\|_2}{\|x_k - x^\dagger\|_2}\right) - \varepsilon\\
\end{eqnarray*}
On the other hand since:
\begin{equation}
    \|(I- \tau A^T{S_i}^TS_iA)(x_k-x^\dagger)\|_2 \leq \|(I- \tau A^T{S_i}^TS_iA)\|_2\|(x_k-x^\dagger)\|_2 \leq (1+\tau qL_s)\|(x_k-x^\dagger)\|_2,
\end{equation}
and also note the second part of restricted eigenvalue condition we have:
\begin{equation}
    \|S_iA(x_k - x^\dagger)\|_2 \leq qL_c \|x_k - x^\dagger\|_2.
\end{equation}
Hence:
\begin{eqnarray*}
&&\|P_{\mc{C}}[(I- \tau A^T{S_i}^TS_iA)(x_k-x^\dagger)]\|_2\\
    &\geq& \|x_k - x^\dagger\|_2\left(1 - \tau\frac{\|S_iA(x_k - x^\dagger)\|_2}{\|x_k - x^\dagger\|_2}\right)  \\
    &\geq&  \left(\frac{1 - q\tau L_c}{1+q\tau L_s}\right) \|(I- \tau A^T{S_i}^TS_iA)(x_k-x^\dagger)\|_2
\end{eqnarray*}
Combining these three with $\tau = \frac{1}{q L_s}$, we find that (\ref{lemo}) is satisfied for the choice $v = (I- \tau A^T{S_i}^TS_iA)(x_k-x^\dagger)$ and $a = \frac{L_s - L_c}{2L_s}$, we can write:
\begin{equation}
    \|x_{k+1}-x^\dagger\|_2 \geq (1-\gamma)(1- \frac{L_c}{L_s}) \|x_{k}-x^\dagger\|_2 - \varepsilon,
\end{equation}
for all $\|x_k - x^\dagger\|_2 \leq \frac{\delta}{2}$ and by unfolding the iterations to $x_0$ we finish the proof.

\section{Proof for Theorem 3.3}

We shall start with the quantifying bounds (Lemma 3.2) of restricted eigenvalues of the construction $A=GB$ where $G \in \mb{R}^{n \times d}$ is a Gaussian map. Applying Gordon's escape-through-a-mesh on Gaussian map, we have:
\begin{equation}\label{b1}
    \|Av\|_2 = \|GBv\|_2 \geq [p_n - \mc{W}(B\mc{C}\cap\mb{S}^{n-1}) - \theta]\|Bv\|_2 \geq \sigma_b[p_n - \mc{W}(B\mc{C}\cap\mb{S}^{n-1}) - \theta]\|v\|_2, \forall v \in \mc{C},
\end{equation}
with probability at least $1 - e^{-\frac{\theta^2}{2}}$. Meanwhile, for a single realization of $S_i$, with probability at least $1 - e^{-\frac{\theta^2}{2}}$ we have:
\begin{equation}\label{b2}
    \|S_iAv\|_2 = \|S_iGBv\|_2 \leq [p_q + \mc{W}(B\mc{C}\cap\mb{S}^{n-1}) + \theta]\|Bv\|_2 \leq \sigma_a[p_n + \mc{W}(B\mc{C}\cap\mb{S}^{n-1}) + \theta]\|v\|_2, \forall v \in \mc{C}.
\end{equation}
For bounding $L_s$, noting that $\mc{W}(B\mc{C}\cap\mb{S}^{n-1}) \leq \sqrt{d}$, we have:
\begin{equation}\label{b3}
    \|S_iAu\|_2 = \|S_iGBu\|_2 \leq \sigma_a[p_n + \sqrt{d} + \theta]\|u\|_2, \forall u \in \mb{R}^d.
\end{equation}
We now turn to the proof of the quantifying upper bound. Following the same steps of the proof for Theorem 3.1, and plugging in (\ref{b1}) and (\ref{b3}), we can have:
\begin{equation}\label{au}
    \alpha = \kappa_\M(1 - \mu_c/L_s) \leq \kappa_\M\left( 1- \frac{q\sigma_b^2 (p_n - \mc{W}(B\mc{C}\cap\mb{S}^{n-1}) - \theta)}{n\sigma_a^2(p_q + \sqrt{d} + \theta)}\right),
\end{equation}
and applying the union bound we know that the above inquality holds with probability at least $1 - m e^{-\frac{\theta^2}{2}}$. Plugging the bound (\ref{au}) to Theorem 3.1 we obtain the upper bound in Theorem 3.3. For the quantifying lower bound:
\begin{equation}
    (1-\gamma)(1- \frac{L_c}{L_s}) \geq (1-\gamma) \left( 1 - \frac{\sigma_b^2(p_q + \mc{W}(B\mc{C}\cap\mb{S}^{n-1}) + \theta)^2}{\sigma_a^2(p_q - \sqrt{d} - \theta)^2} \right)
\end{equation}
with probability at least $1 - m e^{-\frac{\theta^2}{2}}$. Thus finishes the proof.

\section{Unsupervised Training without Ground-truth Data}

Very recently, \cite{chen2021equivariant} proposed a new way of constructing network regularization, named \textit{Equivariant-Imaging} (EI), which utilize the equivariant structure of the imaging systems, and does not require the usage of the ground-truth data. Let $g \in \G$ where $\G$ is a group of transformations (for example, rotation transformation), and we assume that the operation of forward pass of the network $\F_\theta(A\    \cdot)$ is equivariant for this group of transformation:
\begin{equation}
    T_g\F_\theta(Ax) = \F_\theta(AT_g(x)), \ \ \forall x \in X \subset \R^d,
\end{equation}
where $T_g$ is a unitary matrix such that $T_gx \in X, \ \forall x \in X \subset \R^d$. Such a condition is true for CT/PET if we consider rotation operation. Utilizing this structure, the training objective can be written as:
\begin{equation}
\begin{aligned}
    \te^\star \in \arg\min_{\te} \E_{\mu_b, g} [\|b - A(\F_\te(b))\|_2^2 + \lambda_{\mr{EI}} \|T_g\F_\te(b) - \F_\te(AT_g\F_\te(b))\|_2^2].
\end{aligned}
\end{equation}
The network regularization encourages the network to structurally learn the null-space of $A$. Adopting this training procedure in our setting, we can train the our networks only using the measurement data, without the need of ground-truth data.

\section{Additional results on Sparse-view CT}

In this subsection we present some results of our unrolling network for sparse-view CT fan beam reconstruction. We use the ODL toolbox \citep{adler2018learned} to simulate fan beam projection data with 200 equally spaced angles, each has 400 ray. We include a small amount of Posisson noise:
\begin{equation}
    b \sim \mr{Poisson}(I_0 e^{-Ax^\dagger}),
\end{equation}
with $I_0 = 5 \times 10^{6}$. In our network we block-partition (according to the angles) forward operator and data into 4 subsets. The network has 12 layers hence correspond to 3 data-passes, which means it takes only 3 calls in total on the forward and adjoint operator. We compare it with the learned primal-dual which has 12 layers, corresponding to 12 calls on the forward and adjoint operator. We train all the comparing models end-to-end with 20 epochs of ADAM optimizer \citep{kingma2014adam}. Since the subnetworks $\Pm_{\te_p}(\cdot)$ and $\D_{\te_d}(\cdot)$ in each layer are in small scales (3 convolutional layers and 32 channels), the main computational cost came from the forward and adjoint operator. Our network requires only a fraction of the computation (both in training and testing), with little compromise on reconstruction performance, as shown by the results presented in Table 1.

We also present our unsupervised training results of both unrolling networks. Here we choose the group-action for data-augmentation to be rotations with $[90, 180, 270]$ degrees, sampled randomly in each training iteration. We found numerically that under unsupervised training, both networks' performance is degraded compare to supervised setting, while LSPD network result is still competitive with the full batch LPD result, with only a fraction of computation. Note that this is the first result in the literature for unrolling networks trained without ground-truth data.

Moreover, we test the performance of our LSPD network with the instance-adaptation scheme presented in Section 4. We run 50 iteration of ADAM \citep{kingma2014adam} on the instance-adaptation objective for each of the test data sample. We found out that the instance-adaptation can be successfully applied to boost the performance of LSPD network. We present two reconstruction examples in Figure \ref{f3}. Note that the same scheme can be also applied to the full LPD network. However, this is not recommended in practice since for LPD, the instance-adaptation will require much more computation on applying full forward and adjoint operator compare to LSPD, especially in large-scale applications.

\begin{table*}[t]\label{Syn}
\caption{Sparse-View fan-beam CT Result for LPD and LSPD on Mayo dataset}
\label{sample-table1}
\vskip 0.15in
\begin{center}
\begin{small}
\begin{sc}
\begin{tabular}{lcccr}
\hline
Method & $\#$ calls on $A$ and $A^T$& PSNR & SSIM  \\
\hline

&&&\\
FBPConvNet (FBP+UNet denoising) & -- & 33.5975 & 0.8767  \\

LPD \textit{(12 layers)} & 24& 39.3483 & 0.9398  \\

LSPD  \textit{(12 layers)} & 6& 38.1044& 0.9162\\
\\

\textit{Unsupervised training without ground-truth data:}\\
LPD-EI \textit{(12 layers, $\lambda_\mr{EI} = 100$)} & 24& 35.7908& 0.8715  \\

LSPD-EI  \textit{(12 layers, $\lambda_\mr{EI} = 100$)} & 6& 35.0743& 0.8374\\

\hline
\end{tabular}
\end{sc}
\end{small}
\end{center}
\vskip -0.1in
\end{table*}

\begin{figure*}[t]
   \begin{center}

    {\includegraphics[trim=80 80 15
    15,clip,width=0.9\textwidth]{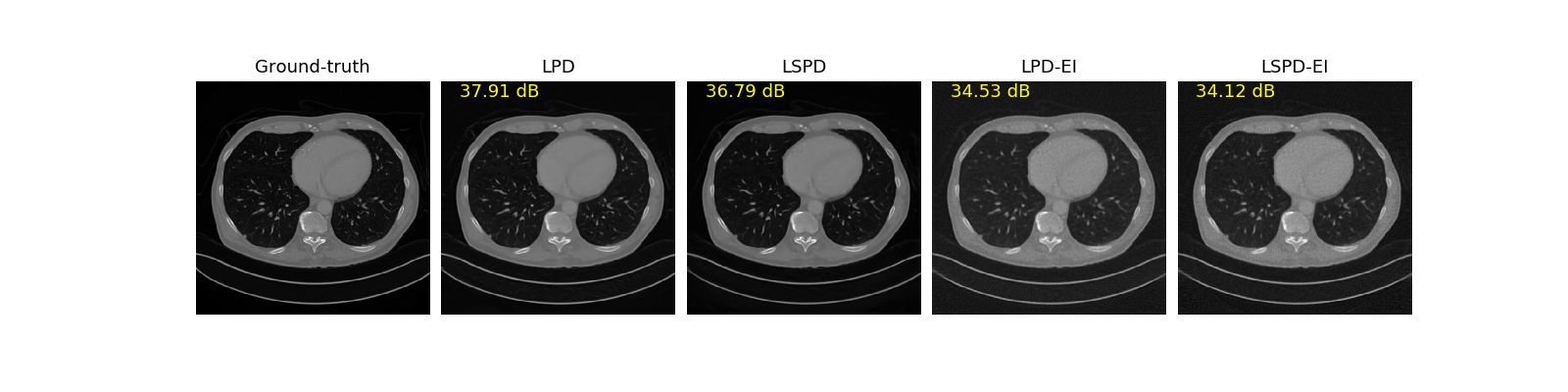}}
    {\includegraphics[trim=80 80 15 55,clip,width=0.9\textwidth]{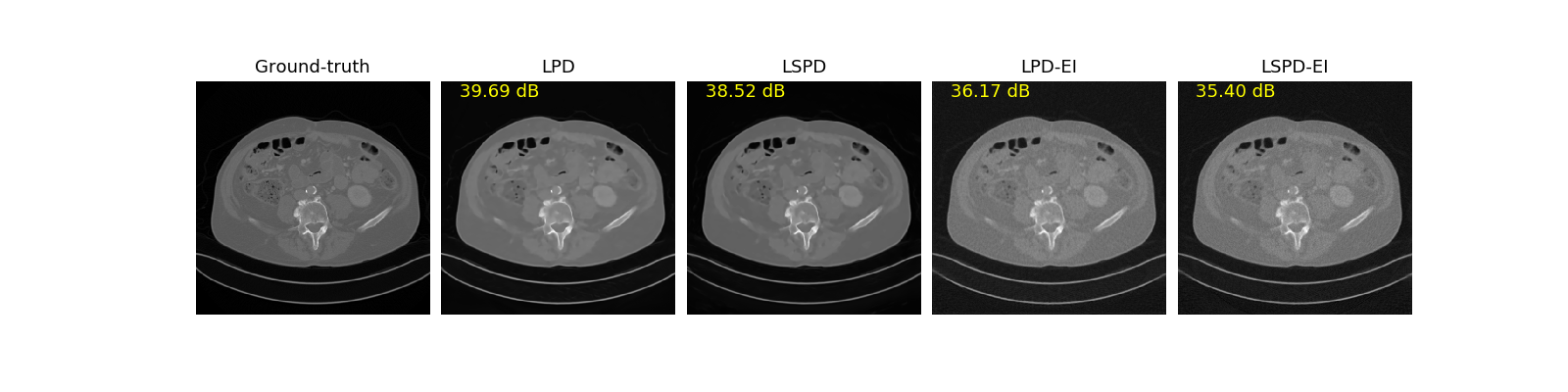}}
        {\includegraphics[trim=80 80 15 55,clip,width=0.9\textwidth]{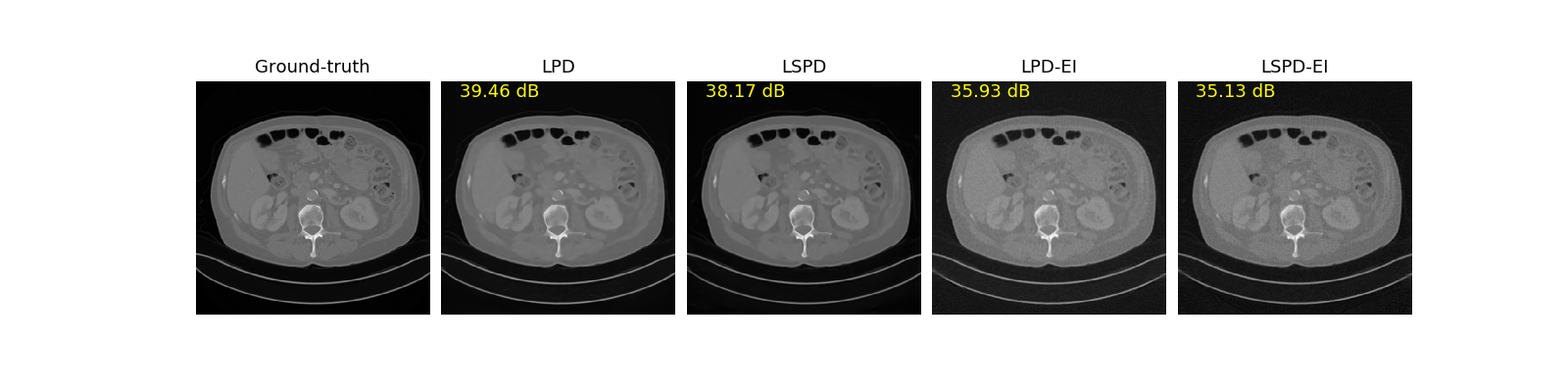}}
    \end{center}
   \caption{Results for Low-dose CT on the test set of Mayo dataset}
    \label{f5}
\end{figure*}

\subsection{Instance-adaptation for out-of-distribution data}

\begin{figure*}[t]
   \begin{center}
    {\includegraphics[trim=80 110 15
    15,clip,width=0.9\textwidth]{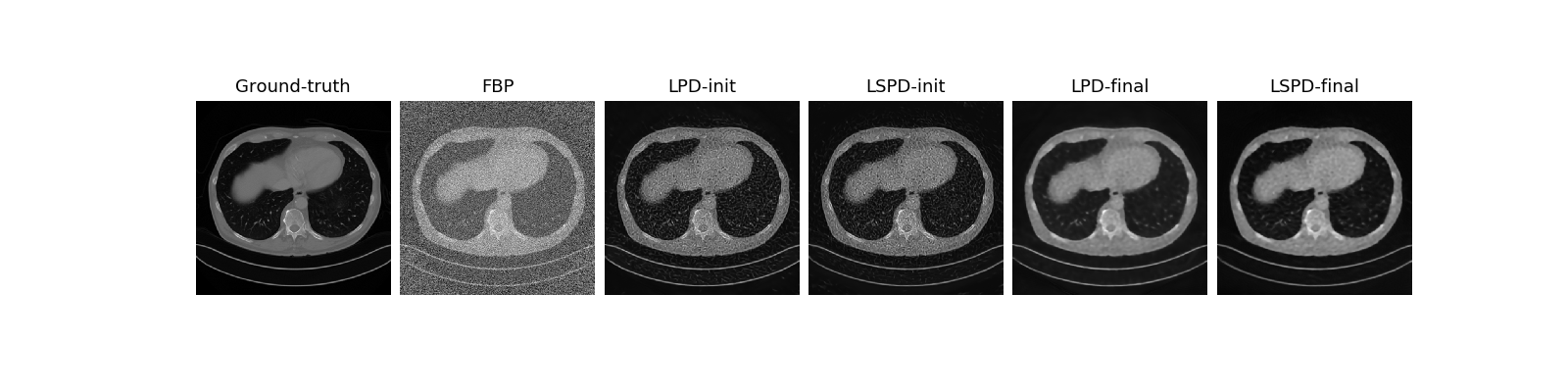}}
        {\includegraphics[trim=80 110 15 115,clip,width=0.9\textwidth]{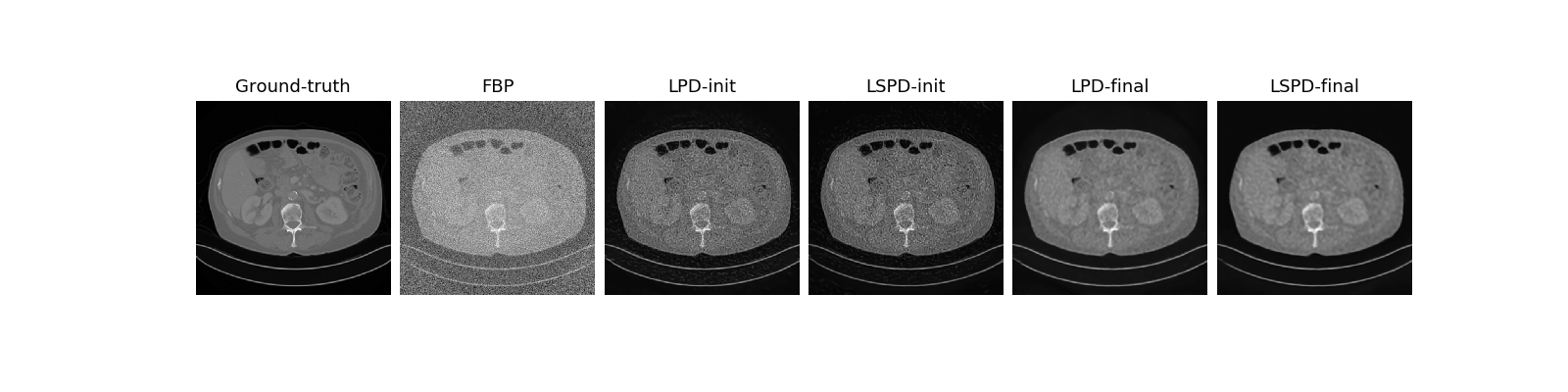}}
    \end{center}
   \caption{Results for instance-adaptation for out-of-distribution reconstruction, with noise level mismatch}
    \label{f6}
\end{figure*}

\begin{figure*}[t]
   \begin{center}
    {\includegraphics[width=0.39\textwidth]{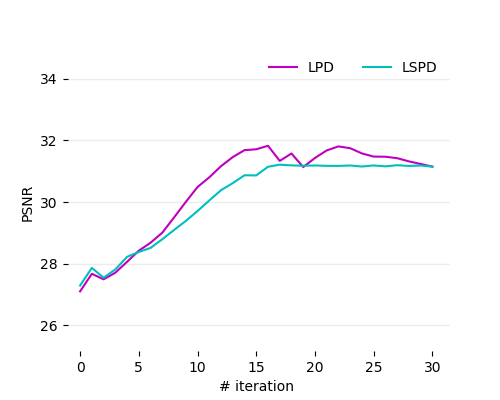}}
    {\includegraphics[width=0.39\textwidth]{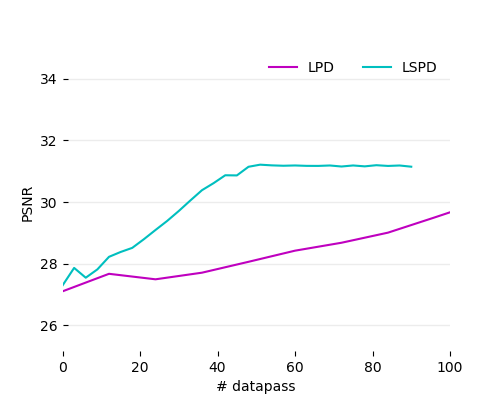}}\\
        {\includegraphics[width=0.39\textwidth]{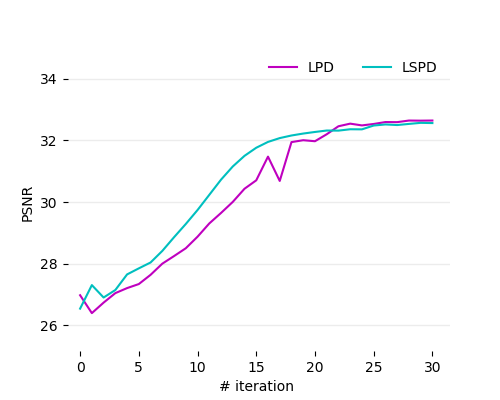}}
    {\includegraphics[width=0.39\textwidth]{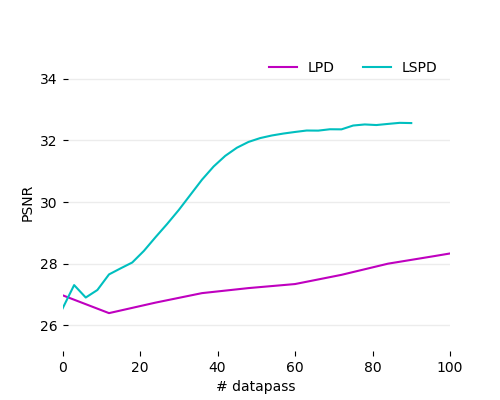}}\\
    \end{center}
   \caption{Results for instance-adaptation of out-of-distribution reconstruction, with noise level mismatch (Row 1 for example 1)}
    \label{f7}
\end{figure*}

We also apply the LPD and LSPD models trained in the previous section to a noisier sparse-view CT task, with $I_0 = 3 \times 10^5$, and test the performance of the models in instance-adaptation. In Figure \ref{f6} we present two slices for illustrative examples. We denote LPD-init and LSPD-int to represent the results of the models trained in previous section which were aimed for a smaller noise level. Due to this mismatch, we can clearly observe a significant degrade on both of the reconstruction. Then we run 30 iterations of Adam \citep{kingma2014adam} on instance-adaptation objective (\ref{ia_ei_obj}) for the input examples. We plot the convergence curve on PSNR in Figure \ref{f7}, showing that our proposed LSPD network can also excel in instance-adaptation tasks. We observe that regarding the convergence rate in terms of the number of iterations, the LPD network has faster initial rate, but slow down in later iterations and catched up by the curve for LSPD, and the LSPD reaches better final recovery results. Noting that for each iteration the LPD requires much more computation than LSPD, hence for a clearer demonstration of the benefit of our LSPD network, we also plot the PSNR against the number of passes of the input data $b_\mathrm{in}$ (the number of calls on $A$ and $A^T$).

\begin{figure*}[t]
   \begin{center}
    {\includegraphics[trim=80 110 15
    15,clip,width=0.9\textwidth]{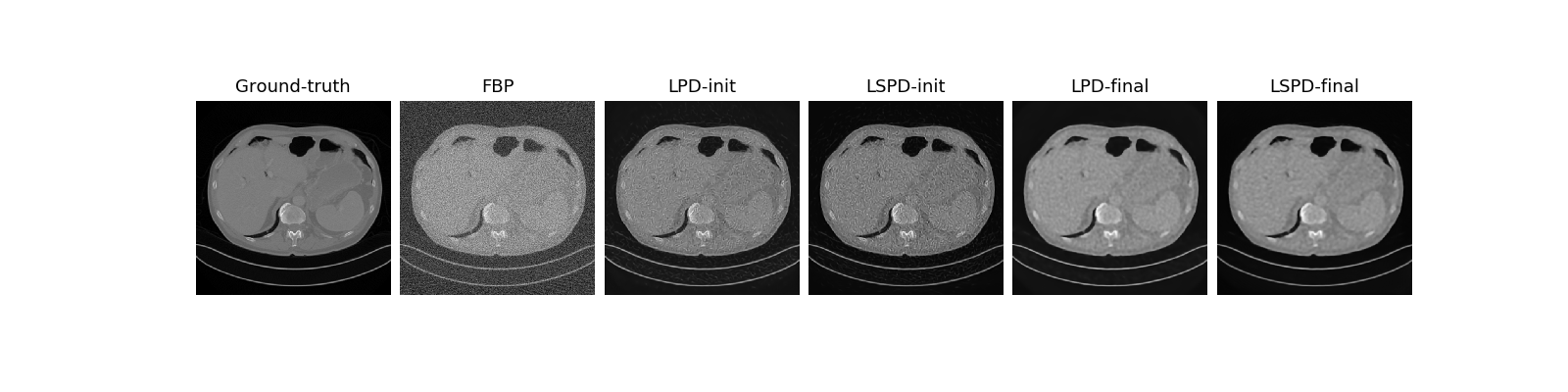}}
        {\includegraphics[trim=80 110 15 115,clip,width=0.9\textwidth]{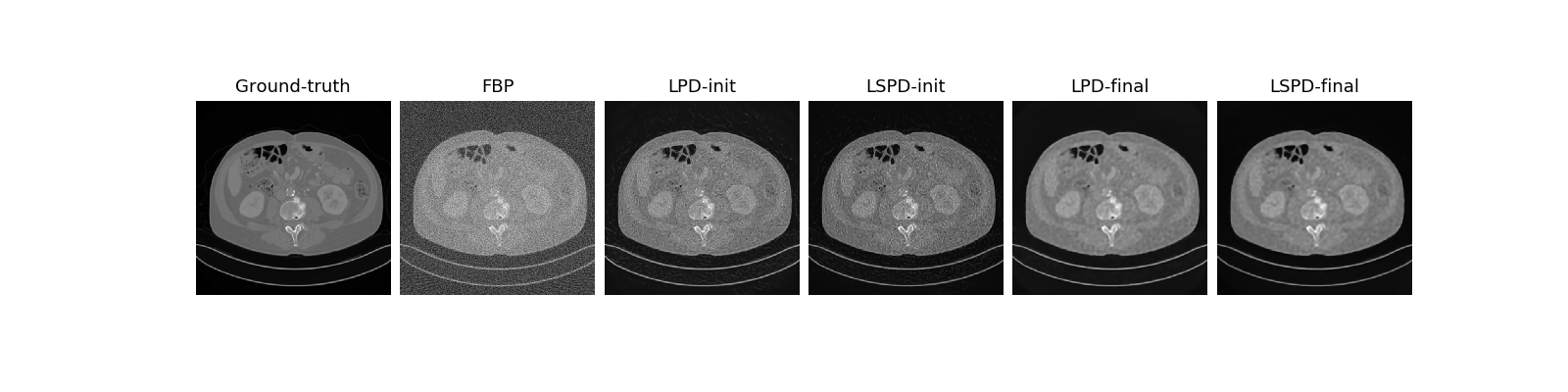}}
    \end{center}
   \caption{Results for instance-adaptation for out-of-distribution reconstruction, with model mismatch}
    \label{f9}
\end{figure*}

\begin{figure*}[t]
   \begin{center}
    {\includegraphics[width=0.39\textwidth]{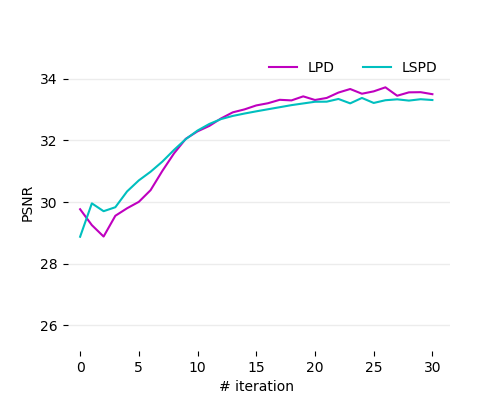}}
    {\includegraphics[width=0.39\textwidth]{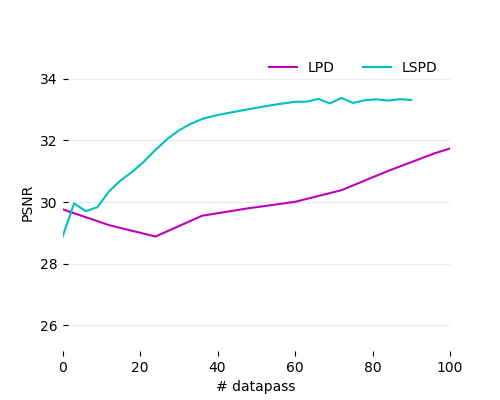}}\\
        {\includegraphics[width=0.39\textwidth]{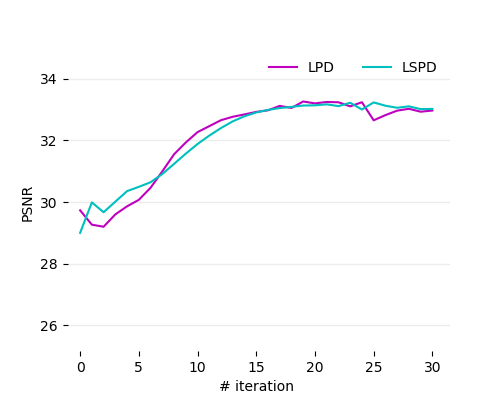}}
    {\includegraphics[width=0.39\textwidth]{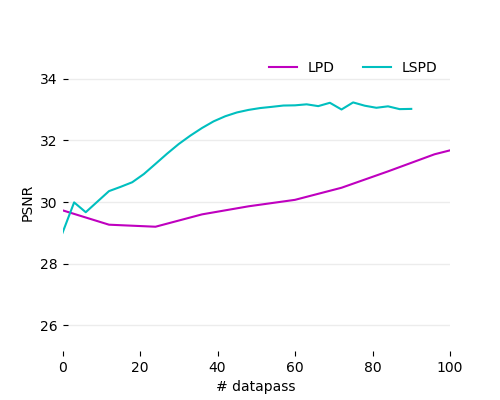}}\\
    \end{center}
   \caption{Results for instance-adaptation of out-of-distribution reconstruction, with model mismatch (Row 1 for example 1)}
    \label{f10}
\end{figure*}

Meanwhile, we also evaluate the instance-adaptation performance of both networks at the presence of model mismatch (Figure \ref{f9} and \ref{f10}), where the testing data is obtained from a different scanner geometry than the training data, with a doubled X-ray source distance and $I_0 = 5 \times 10^5$. We again observe a much improved performance in terms of both convergence rate and recovery accuracy for our proposed network LSPD over the classical LPD network.

\begin{figure*}[t]
   \begin{center}
    {\includegraphics[trim=80 110 15
    15,clip,width=0.9\textwidth]{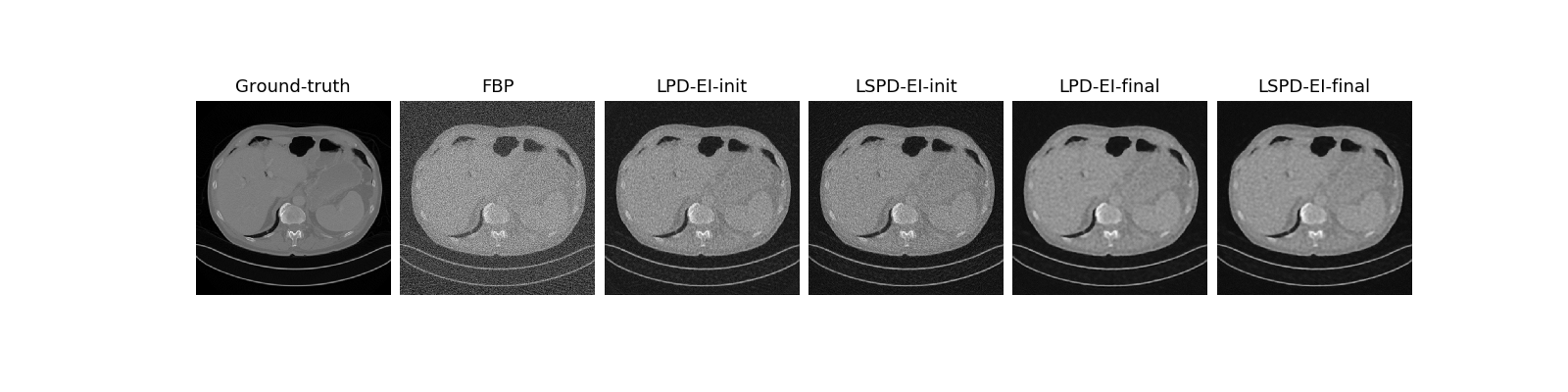}}
        {\includegraphics[trim=80 110 15 115,clip,width=0.9\textwidth]{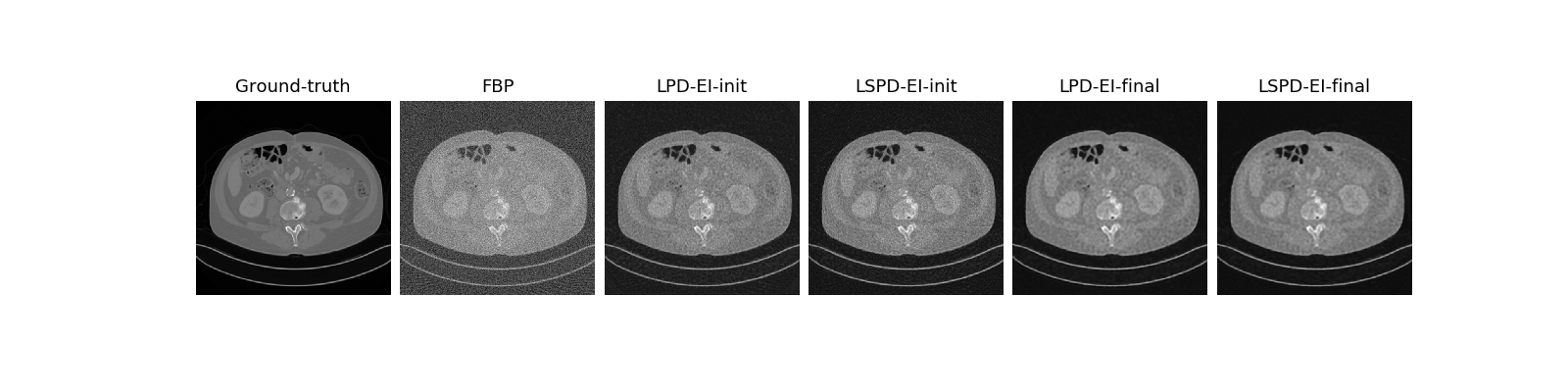}}
    \end{center}
   \caption{Results for instance-adaptation for out-of-distribution reconstruction where the models are pretrained without ground-truth), with model mismatch}
    \label{f11}
\end{figure*}

\begin{figure*}[t]
   \begin{center}
    {\includegraphics[width=0.39\textwidth]{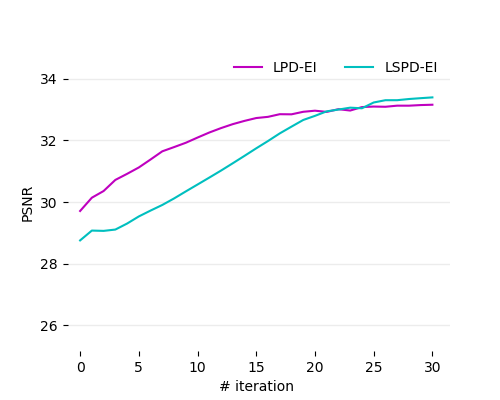}}
    {\includegraphics[width=0.39\textwidth]{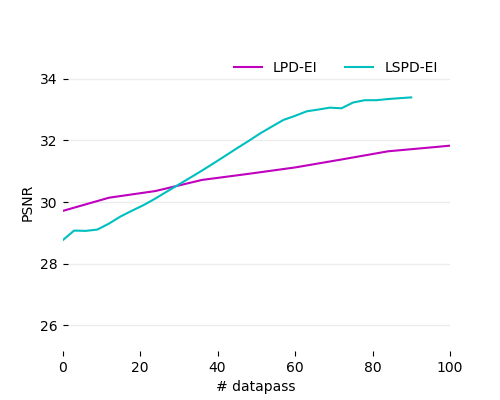}}\\
        {\includegraphics[width=0.39\textwidth]{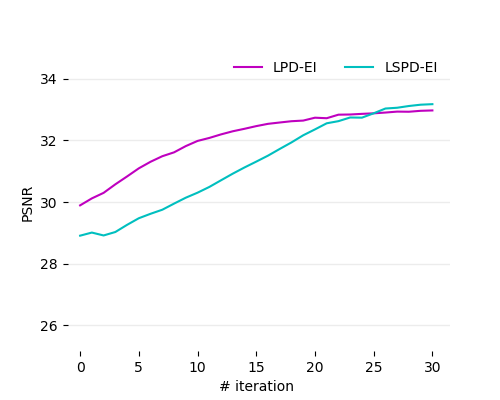}}
    {\includegraphics[width=0.39\textwidth]{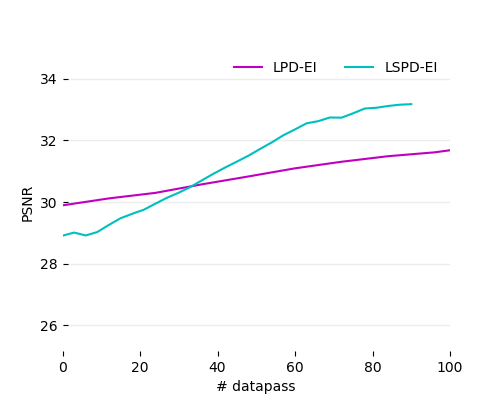}}\\
    \end{center}
   \caption{Results for instance-adaptation of out-of-distribution reconstruction (where the models are pretrained without ground-truth), with model mismatch (Row 1 for example 1)}
    \label{f12}
\end{figure*}

\end{document}